\documentclass[12pt]{iopart}

\usepackage{iopams}
\newtheorem{lemma}{Lemma}
\newtheorem{proposition}{Proposition}
\newtheorem{theorem}{Theorem}

\newcommand{\be}{\begin{equation}}
\newcommand{\ee}{\end{equation}}
\def\tr{\mathop{\rm tr}\nolimits}

\def\dif{{\rm d}}

\def\Riem{{\rm Riem}}

\def\r{{\rm r}}

\def\R{{\rm R}}

\begin{document}

\title[Homogeneous three-dimensional Riemannian spaces]{Homogeneous three-dimensional Riemannian spaces}

\author{Joan Josep Ferrando$^{1,2}$ and  Juan Antonio S\'aez$^3$}

\address{$^1$\ Departament d'Astronomia i Astrof\'{\i}sica, Universitat
de Val\`encia, E-46100 Burjassot, Val\`encia, Spain}

\address{$^2$\ Observatori Astron\`omic, Universitat
de Val\`encia, E-46980 Paterna, Val\`encia, Spain}

\address{$^3$\ Departament de Matem\`atiques per a l'Economia i l'Empresa,
Universitat de Val\`encia, E-46022 Val\`encia, Spain}

\ead{joan.ferrando@uv.es; juan.a.saez@uv.es}
\begin{abstract}
The necessary and sufficient conditions for a three-dimensional Riemannian metric to admit a transitive group of isometries are obtained. These conditions are Intrinsic, Deductive, Explicit and ALgorithmic, and they offer an IDEAL labeling of these geometries. It is shown that the transitive action of the group naturally falls into an unfolding of some of the ten types in the Bianchi-Behr classification. Explicit conditions, depending on the Ricci tensor, are obtained that characterize all these types.
\end{abstract}
\pacs{04.20.-q, 02.20.Sv, 02.40.Ky}
%
%

\submitto{\CQG}


\section{Introduction}
\label{sec-intro}

The spatially homogeneous cosmological models (Bianchi models) generalize the Friedmann-Lema\^itre-Robertson-Walker solutions. Nevertheless, although the Einstein equations also become an ordinary differential system, they are compatible with anisotropies produced by rotation or global magnetic fields, and thus they can be more suitable to model the real universe, or to approximate it in some regions. 

There is ample literature on Bianchi models dating back to the fifties \cite{taub, schucking} after which research took a crucial step forward in the late sixties \cite{E-W-Behr, Ellis-McCallum} with the use of the orthonormal frame formalism. Ryan-Shepley's book \cite{RyanShepley} offers an exhaustive bibliography before 1975, and several more recent manuals present the basic concepts and references on the subject   
 \cite{Ellis-Elst, Kramer, Krasinski-Plebanski, ellis-maar-mac}. The study of new solutions and the analysis of the geometric properties of the Bianchi models are currently ongoing (see the recent paper \cite{thorsrud} and references therein).
 
A question that remains open on the spatially homogeneous solutions is its IDEAL characterization, that is, its labeling through conditions that are Intrinsic (depending only on the
metric tensor), Deductive (not involving inductive or
inferential methods or arguments), Explicit (the solutions
are not expressed implicitly) and Algorithmic (giving the
solution as a flow chart with a finite number of steps). 

In this paper we present the first necessary step for implementing this approach: the IDEAL labeling of the homogeneous three-dimensional Riemannian spaces. We make use of the invariant approach by Bona and Coll \cite{bonacoll1, bonacoll2}, which gives the necessary and sufficient conditions for a three-dimensional Riemannian metric to admit a group G$_r$ of isometries acting on s-dimensional orbits. Their study is certainly invariant because the conditions are expressed in terms of the eigenvalues and eigenvectors of the Ricci tensor. Nevertheless, some additional work is necessary to acquire a fully IDEAL characterization and to discriminate the different Bianchi types.  

The Bianchi method \cite{bianchi} to classify the three-dimensional Lie algebras was based on the Lie groups theory and it leads to the original nine Bianchi types. Nowadays, the commonly considered classification follows the Sch\"{u}cking-Kundt-Behr approach \cite{E-W-Behr, Ellis-McCallum, Kramer, Krasinski-Plebanski, kundt}. This method consists in defining, in terms of the structure constants, two constant matrices $A_a$ and $n^{ab}$, and characterizing the types by considering the properties which are independent of the representation of the algebra: the rank and signature of $n^{ab}$ and the vanishing or not of $A_a$. This procedure leads to ten classes (in what follows Bianchi-Behr types)
that basically coincide with the Bianchi types.

It is worth remarking that the Bianchi and the Bianchi-Behr types concern a three-dimensional abstract Lie algebra with no reference to the space where the Lie group acts. Nevertheless, we are interested in a transitive action on a three-dimensional Riemannian space. This means that we are actually analyzing and classifying Riemannian metrics. From this point of view a sub-classification appears in a natural way. In this paper we analyze this fact and offer an IDEAL labelling of all these classes of Riemannian metrics.

In section \ref{sec-SKB} we introduce {\em the structure tensor of a group in simply-transitive action}. This structure tensor $Z$ defines, and is defined by, a vector $a$ and a symmetric tensor $N$. They are the tensorial version of the matrices $A_a$ and $n^{ab}$ determining the Bianchi-Behr classification. The action of the group on the Riemannian space facilitates this tensorial approach, which is necessary to obtain the IDEAL labeling of the metrics admitting each Bianchi-Behr type in next sections. We also present in this section an algorithm, based on explicit algebraic functions depending on $N$ and $a$, which distinguishes the Bianchi-Behr types once the structure tensor $Z$ is given. 

Section \ref{sec-multiply} is devoted to studying the multiply-transitive actions. By starting from the results by Bona and Coll \cite{bonacoll1, bonacoll2} we offer an explicit and invariant characterization of the three classes admitting a G$_6$ and the three classes admitting a G$_4$. The conditions involved are algebraic in the Ricci tensor for the G$_6$ case and differential of first-order for G$_4$. 

In section \ref{sec-simply} we consider the simply-transitive action of a G$_3$ and we also obtain an explicit and intrinsic labeling of the metrics, with the structure tensor $Z$ playing an important role. The structure tensor $Z$ can be obtained from the Ricci tensor and its first derivatives when the Ricci tensor is algebraically general. When the Ricci tensor is algebraically special $Z$ also depends on second derivatives of the Ricci tensor. And then, the conditions involved in the characterization include, at most, derivatives of third-order. We also summarize the results in the last two sections by presenting a flow chart enabling us to distinguish in algorithmic form the dimension of the groups acting transitively in a three-dimensional Riemannian space.

The study and characterization of the transitive action of a specific Bianchi-Behr type is tackled in section \ref{sec-ideal-BB}. We recover some results about the non-maximal action of the Bianchi types I, II, III and V, and we study the other Bianchi-Behr types that act transitively when the maximal group is a G$_4$ or a G$_6$. In the study of the maximal action of the Bianchi-Behr types we must distinguish the metrics with an algebraically special Ricci tensor separately, and the simple eigenvalue and the rotation of the simple eigenvector distinguish the three possible types. When the Ricci tensor is algebraically general, we use the expression of the structure tensor in terms of the Ricci tensor and we apply the algorithm presented is section \ref{sec-SKB} to label the compatible Bianchi-Behr types. In this section we also present several diagrams that emphasize the algorithmic nature of the results.

In section \ref{sec-comments} we remark the conceptual and practical interest of an IDEAL characterization and we comment on our results and on the work in progress on this subject.

\ref{A-Ricci} presents the algebraic study of the Ricci tensor for a Riemannian space when an isometry group of a specific Bianchi-Behr type acts. In \ref{A-AS} we study the kinematic properties of the eigenvector associated with the simple eigenvalue of an algebraically special Ricci tensor.


\section{The Sch\"{u}cking-Kundt-Behr method in tensorial formalism}
\label{sec-SKB}

Let $g$ be a three-dimensional Riemannian metric admitting a transitive group $G_3$ of isometries, and let $\{ \xi_a \}$ be an oriented basis defined by three independent Killing vector fields. If $|g|$ is the determinant of the metric in this basis, $|g| = det(g_{ab})$, $g_{ab}= g(\xi_a, \xi_b)$, $\{\theta^a\}$ is the algebraic dual basis and $\eta$ is the metric volume element, we have  
\begin{equation} \label{epsilon}
\eta = \frac{1}{\sqrt{|g|}} \ \xi_1 \wedge \xi_2 \wedge  \xi_3 \, , \qquad  \theta^a =  \frac{\epsilon^{abc}}{2 \sqrt{|g|}} *( \xi_b \wedge \xi_c)   \, .
\end{equation}
where $*$ denotes the Hodge dual operator. Moreover, $[\xi_a, \xi_b] = C_{\, ab}^{c}\, \xi_c$, where $C_{\, ab}^{c}$ are the structure constants. In what follows we denote with the same symbol any tensor and its associated ones by raising and lowering indexes with the metric tensor.  

As $\xi_a$ is a Killing field, its covariant derivative $\nabla \xi_a $ is a 2-form, which is determined by its dual vector $\phi_a$ in a three dimensional space. Then, associated with the basis $\{ \xi_a \}$, we can define the tensor
\begin{equation} \label{zeta}
Z= \theta^a \otimes \phi_a \, ,\qquad  \phi_a \equiv *\nabla \xi_a \, .
\end{equation}

Note that the tensor $Z$ is independent of the representation of the Lie algebra,
that is, it is invariant under a linear transformation $\{ K_b^a \xi_a \}$ that preserves orientation ($|(K_a^b)|>0$). We call $Z$ the {\em structure tensor} of a group in simply-transitive action. Moreover, $Z$ satisfies
\begin{equation} \label{siete}
\phi_a = i(\xi_a)  Z \, , \qquad \nabla \xi_a = *i(\xi_a)  Z \, . 
\end{equation}
where $i(\xi)t$ means the interior product of a vector field $\xi$ with a p-tensor $t$. 

On the other hand, we obtain ${\cal L}_{\xi_a} \theta^b = - C_{\, ac}^b \theta^c$ and ${\cal L}_{\xi_a} \phi_b =  C_{\, ab}^c \phi_c$, and then one can easily show that $Z$ is invariant under the isometry group, ${\cal L}_{\xi_a}Z =0$. This property is equivalent to the following expression that gives the covariant derivative $\nabla Z$ in terms of $Z$:
\be \label{nablaZ}
\nabla_{k} Z_{ij} = Z_k^{\ m} (\eta_{i m n} Z^n_{\ j} +  \eta_{j m n} Z_{i}^{\ n} )  \,  .
\ee

From the Killing integrability conditions for the three independent Killing fields $\xi_a$, $\nabla \nabla
\xi_a = i(\xi_a) \Riem $, and taking into account (\ref{siete}) and (\ref{nablaZ}), we can obtain the Riemann tensor as:
\begin{equation} \label{riemann}
R_{ij}^{\ \, kl} = ({{Z_{i}}^{\, m}} \, {\eta_{m j n}} \, {Z}^{n}_{\ p}  + \nabla_j Z_{i p}) \, \eta^{p k l}  = 
\eta_{m n [i} \, Z_{j]}^{\ m} \, Z^{n}_{\ p} \, \eta^{p k l} - Z_{[i}^{\ k}  \, Z_{j]}^{\ l}   \, ,
\end{equation}
where $[ij]$ denotes antisymmetrization.

The structure tensor $Z$ is a general 2-tensor which can
be decomposed in terms of a symmetric tensor $N$ and a vector $a$,
\begin{equation} \label{eseya}
\hspace{-10mm} Z = N - \frac12 (\tr N) g - *a \, , \quad   N_{ij}= Z_{(ij)} -  (\tr Z) g_{ij} \, ,
\quad  a = - * Z \, .
\end{equation}
where $(ij)$ denotes symmetrization.

From now on, for two 2-tensors $A$ and $B$, $A \cdot B$ denotes the 2-tensor $(A \cdot B)_{ij} = A_i^{\ k} B_{kj}$, and $A^2 = A \cdot A$, $A^3=A^2 \cdot A$. Now, we introduce the following notation:
\be \label{N-scalars}
\nu \equiv \tr N \, , \qquad \mu \equiv \tr N^2 \, , \qquad \lambda \equiv \tr N^3 \, .
\ee

From the decomposition (\ref{eseya}) and the first expression in (\ref{riemann}) we obtain the Ricci tensor $R$ as:
\begin{equation} \label{ricci}
R = 2 N^2 -  \nu N  +\Big[ \frac12 \nu^2 - \mu \Big] \, g - 2 \, a \otimes a +2 \, \nabla a \, .
\end{equation}

From this expression and the symmetry of the Ricci tensor it follows that $a$ is a closed 1-form. In fact, the potential of $a$ depends of the metric determinant as a consequence of (\ref{siete}) and (\ref{eseya}). More precisely, we obtain:
\begin{equation}  \label{da}
\dif a =0 \, , \qquad   \dif \ln |g| = - 4 \, a \, .
\end{equation}
Consequently, {\em the necessary and sufficient condition for $Z$ to be a symmetric
tensor ($a=0$) is that the determinant of the metric $|g|$ be a constant}.

On the other hand, the second expression in (\ref{riemann}) for the Riemann tensor allows us to obtain an
expression for the Ricci tensor which does not involve the derivative of $a$:
\begin{equation} \label{ricci2}
R = 2 N^2 -  \nu N  +\Big( \frac12 \nu^2 - \mu - 2 a^2\Big) \, g - (2 N- \nu g) \cdot *a  \, .
\end{equation}
Moreover, taking into account the symmetric and antisymmetric parts of (\ref{nablaZ}), the following constrains between $R$, $N$ and $a$ can be deduced:
\be
N(a) = 0 \, , \qquad R(a) = \Big( \frac12 \nu^2 - \mu - 2 a^2\Big) a \, , \qquad  \nabla \cdot N = \nu \, a \, .
\ee


\subsection{The Bianchi-Behr types}
\label{subsec-BB}

We can compute the components of the vector $a$ and the 2-tensor $N$ in the basis $\{ \xi_a \}$ in terms of the structure constants $C_{\, ab}^{c}$ and we obtain:
\be \label{N-a-C}
\hspace{-15mm} N = \frac{1}{\sqrt{|g|}} n^{ab} \xi_a \otimes \xi_b  \, , \quad a = A_a \theta^a \, , \quad n^{ab} \equiv \frac12 \epsilon^{mn[a} C_{\, mn}^{b]}  \, , \quad A_a \equiv \frac12 C_{\, ba}^{b} \, . 
\ee
If we consider the usual representation of the Lie algebra (see for example \cite{petrov} \cite{kundt}) presented in table \ref{table-1} we obtain, for each Bianchi type, the expressions of $N$ and $a$ given in the last two columns of table \ref{table-1}.

%
%
\begin{table}[t]
$$\begin{array}{l||c|c|c|c|c||}
 & [\xi_1, \xi_2 ]& [\xi_2, \xi_3 ]& [\xi_3, \xi_1]&{\sqrt{|g|}}\  N & {\sqrt{|g|}}\ a
\\ \hline \hline \hline
 I &0 & 0 & 0 & 0 & 0 \\ \hline
II &0 &  \xi_1 & 0 &  \xi_1 \otimes \xi_1 &0 \\ \hline III & 0 & 0 &
-\xi_1 &
 -\frac{1}{2} \ \xi_1
\widetilde{\otimes} \xi_2   & \frac{1}{2} *(\xi_1\wedge \xi_2 )
\\ \hline

IV & 0   & \xi_1 + \xi_2 &   -  \xi_1 &  \xi_1 \otimes \xi_1 &
*(\xi_1\wedge \xi_2 ) \\ \hline

V &  0  &\xi_2 &  -  \xi_1 & 0 & *(\xi_1\wedge \xi_2 ) \\ \hline

VI &   0  &  q   \xi_2 & - \xi_1 & \frac{(q-1)}{2} \  \xi_1
\widetilde{\otimes} \xi_2 & \ \frac{q+1}{2} *(\xi_1\wedge \xi_2 ) \
\\ \hline

VII &   0  &  - \xi_1+ q \xi_2 &   -\xi_2  & \ \frac{q}{2} \xi_1
\widetilde{\otimes} \xi_2
  -   \xi_1 \otimes \xi_1 -  \xi_2 \otimes \xi_2 \  & \ \frac{q}{2}   *(\xi_1\wedge \xi_2 ) \
  \\ \hline
VIII &   \xi_1  & \xi_3   & - 2 \xi_2 &  \xi_1
\widetilde{\otimes}\xi_3  - 2 \xi_2 \otimes \xi_2 & 0 \\
\hline IX & \xi_3  &  \xi_1 &  \xi_2 & \xi_1
 \otimes \xi_1 + \xi_2 \otimes \xi_2 + \xi_3 \otimes \xi_3  & 0 \\
 \hline \hline \hline
\end{array}$$
\caption{For each Bianchi type, this table provides the expression of the commutators in the usual canonical representation of the Lie algebra and the expression of the vector $a$ and 2-tensor $N$. The Bianchi type VI depends on the parameter $q \not= 0, 1$. The Bianchi type VII depends on the parameter $q$, $q^2 < 4$.}
\label{table-1}
\end{table}

Note that the constant matrices $n^{ab}$ and $A_a$ defined in (\ref{N-a-C}) are those used in the Sch\"{u}cking-Kundt-Behr \cite{E-W-Behr, Ellis-McCallum, kundt, Kramer, Krasinski-Plebanski} method to characterize the Bianchi types. Thus, we can recover this classification by considering the nullity or not of the vector $a$ and the rank and signature of the 2-tensor $N$. Indeed, taking into account the expressions in table \ref{table-1} and with an elementary reasoning we find again the ten types of the Sch\"{u}cking-Kundt-Behr classification  \cite{E-W-Behr, Ellis-McCallum} (see table \ref{table-2}). 
%
%
\begin{table}[h]
$$\begin{array}{l||c|c|c|c|c|c||}
 & \quad \ 0[0] \ \quad & \quad \ 1[1] \quad \ & \ \quad  2[0]  \quad \ &\  \quad  2[2]  \quad  \ & \ \quad  3[1]  \quad \ & \ \quad 3[3] \quad \ \\ \hline \hline \hline
\quad  a=0 \quad & I & II & VI_0 & VII_0 & VIII & IX  \\ \hline
\quad a \not=0 \quad  & V &  IV & VI_h (III)  &  VII_h & --- & ---  
\\ \hline

 \hline \hline \hline
\end{array}$$
\caption{This cross-table provides the ten Bianchi-Behr types. The first row presents all the available values of the pairs $r_{N}[s_{N}]$, that give the rank $r_N$ and the signature $s_N$ of the tensor $N$. The Sch\"{u}cking-Kundt-Behr classification also distinguishes whether the vector $a$ is zero or not. The case $a \not=0$ is not compatible with $r_N=3$ because $N(a)=0$. The invariant parameter $h$ is related with the parameter $q$ in table \ref{table-1} by: $h \equiv - \frac{(q+1)^2}{(q-1)^2} > 0$, in Bianchi type $VI$; and $h \equiv  \frac{q^2 }{4-q^2} < 0$ in Bianchi type $VII$. In this classification, the Bianchi type III corresponds to the type $VI_h$ with $h=-1$.}
\label{table-2}
\end{table}


\subsection{The Bianchi-Behr types in algorithmic form}
\label{subsec-algo-BB}

The conditions that fix the rank and signature of the 2-tensor $N$ can be easily stated in terms of the scalar invariants associated with $N$ given in (\ref{N-scalars}). Let us define:
\be \label{nus}
\nu \equiv \tr N \, , \qquad  \nu_2 \equiv \frac12 (\nu^2- \mu) \, , \qquad \nu_3 \equiv \frac13 \lambda - \frac12 \nu \mu + \frac16 \nu^3 \, .
\ee
It is easy to show that $N=0$ if, and only if, the three invariants above vanish; if $\nu \not=0$ and $\nu_2 = \nu_3 =0$, then $r_N=1$; if $\nu \not=0 \not= \nu_2$ and $\nu_3 =0$, then $r_N=2$, and the sign of $\nu_2$ determines the signature $s_N$; otherwise the rank is $r_N = 3$, and the signature is $s_N = 3$ if, and only if, $\nu \nu_3 >0$ and $\nu_2 >0$. All these results allow us to build a flow diagram that offers an algorithm distinguishing the Bianchi-Behr types. In this algorithm we use the three scalars $\nu$, $\nu_2$ and $\nu_3$ and the vector $a$ as initial data. The parameter $h$ of the Bianchi types $VI_h$ and $VII_h$ is the invariant $h = \frac{a^2}{\nu_2}$.


The algorithm in figure \ref{figure-1} is based on explicit algebraic functions depending on $N$ and $a$, that is, on the structure tensor $Z$. Thus, it offers a tensorial version of the Sch\"{u}cking-Kundt-Behr method to label the Bianchi-Behr types. It is worth remarking that it provides two improvements. On one hand, the involved conditions are explicit in the structure tensor $Z$. On the other hand, working in tensorial formalism will allow us to give these conditions in terms of the Ricci tensor in next sections and, consequently, to obtain an IDEAL characterization. In fact, we will obtain the structure tensor as a concomitant of the Ricci tensor when the action of the isometry group is simply-transitive.

%
%
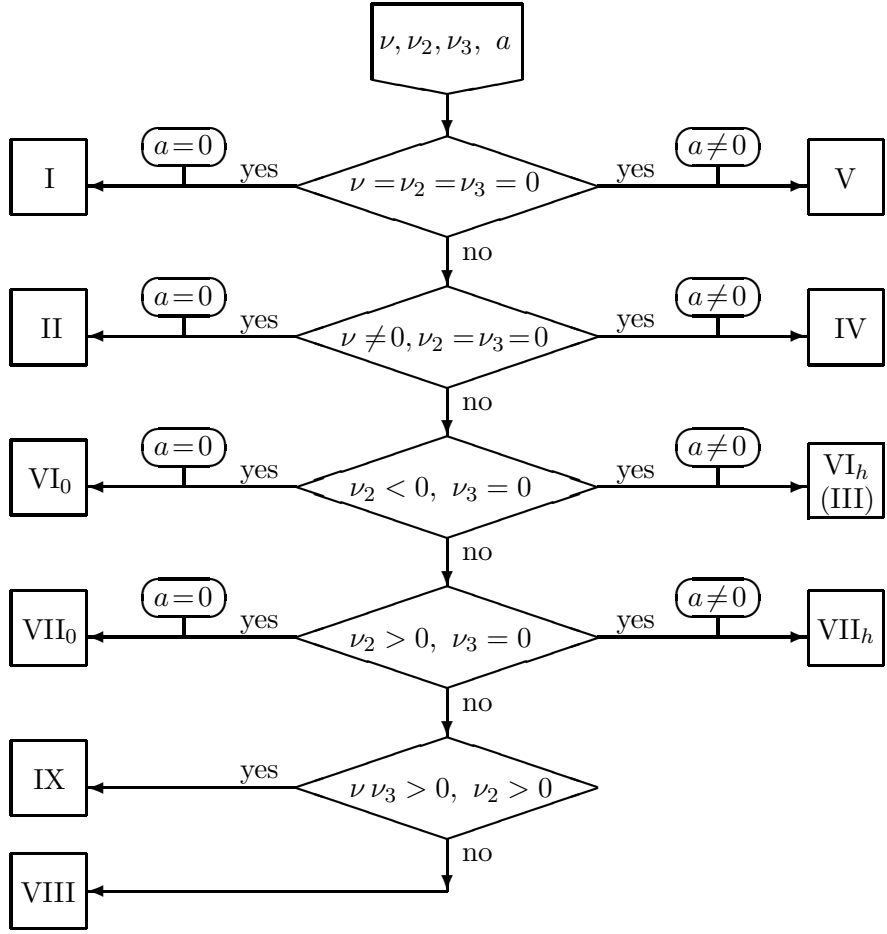
\begin{figure} \label{figure-1}
\begin{center}
\setlength{\unitlength}{1cm} {\small \noindent
\begin{picture}(8,18)
\thicklines
\put(5,17){\line(-5,-1){1}}
 \put(3,17){\line(5,-1){1}}
\put(3,17){\line(0,1){1}} \put(5,18){\line(-1,0){2}}
\put(5,18){\line(0,-1){1}} \put(3.1,17.4){$\nu, \nu_{2},  \nu_3  , \
a $}

\put(4,16.8){\vector(0,-1){0.55}}


\put(6,15.58){\vector(1,0){2.78}} \put(2,15.58){\vector(-1,0){2.78}}


\put(4,16.25){\line(-3,-1){2}}  \put(4,16.25){\line(3,-1){2}}
\put(4,14.9){\line(3,1){2}} \put(4,14.9){\line(-3,1){2}}
\put(2.7,15.51){$\nu=\!\nu_2 =\! \nu_3 =0$}


\put(0.1,16){$a\!=\!0$} \put(1.25,15.7){yes}

\put(0.5,16.1){\oval(1.1,0.5)} \put(0.5,15.84){\line(0,-1){0.25}}


\put(7.2,16){$a\!\neq\!0$} \put(6.25,15.7){yes}

\put(7.6,16.1){\oval(1.1,0.5)} \put(7.6,15.84){\line(0,-1){0.25}}

\put(8.8,15.2){\line(1,0){1}} \put(8.8,15.2){\line(0,1){1}}
\put(9.8,16.2){\line(-1,0){1}} \put(9.8,16.2){\line(0,-1){1}}

\put(9.15,15.56){V}

\put(-1.8,15.2){\line(1,0){1}} \put(-1.8,15.2){\line(0,1){1}}
\put(-0.8,16.2){\line(-1,0){1}} \put(-0.8,16.2){\line(0,-1){1}}

\put(-1.35,15.56){I}

\put(4,14.9){\vector(0,-1){0.65}}

\put(4.2,14.6){no}



\put(4,14.25){\line(-3,-1){2}}  \put(4,14.25){\line(3,-1){2}}
\put(4,12.9){\line(3,1){2}} \put(4,12.9){\line(-3,1){2}}
\put(2.58,13.45){$\nu\neq\! 0, \nu_2 = \! \nu_3\! =\!0$}



\put(0.1,14){$a\!=\!0$} \put(1.25,13.7){yes}

\put(0.5,14.1){\oval(1.1,0.5)} \put(0.5,13.84){\line(0,-1){0.25}}


\put(7.2,14){$a\!\neq\!0$} \put(6.25,13.7){yes}

\put(7.6,14.1){\oval(1.1,0.5)} \put(7.6,13.84){\line(0,-1){0.25}}


\put(6,13.58){\vector(1,0){2.78}} \put(2,13.58){\vector(-1,0){2.78}}

\put(8.8,13.2){\line(1,0){1}} \put(8.8,13.2){\line(0,1){1}}
\put(9.8,14.2){\line(-1,0){1}} \put(9.8,14.2){\line(0,-1){1}}

\put(9.15,13.56){IV}

\put(-1.8,13.2){\line(1,0){1}} \put(-1.8,13.2){\line(0,1){1}}
\put(-0.8,14.2){\line(-1,0){1}} \put(-0.8,14.2){\line(0,-1){1}}

\put(-1.4,13.56){II}


\put(4,12.25){\line(-3,-1){2}}  \put(4,12.25){\line(3,-1){2}}
\put(4,10.9){\line(3,1){2}} \put(4,10.9){\line(-3,1){2}}
\put(2.7,11.45){$ \nu_2 <0, \  \nu_3 =0$}

\put(4,12.9){\vector(0,-1){0.65}}

\put(4.2,12.6){no}


\put(0.1,12){$a\!=\!0$} \put(1.25,11.7){yes}

\put(0.5,12.1){\oval(1.1,0.5)} \put(0.5,11.84){\line(0,-1){0.25}}


\put(7.2,12){$a\!\neq\!0$} \put(6.25,11.7){yes}

\put(7.6,12.1){\oval(1.1,0.5)} \put(7.6,11.84){\line(0,-1){0.25}}


\put(6,11.58){\vector(1,0){2.78}} \put(2,11.58){\vector(-1,0){2.78}}

\put(8.8,11.17){\line(1,0){1}}

\put(8.8,11.15){\line(0,1){1}}

\put(9.8,12.15){\line(-1,0){1}}

\put(9.8,12.17){\line(0,-1){1}}


\put(9,11.75){VI$_h\!$ } \put(8.95,11.30){(III) }


\put(-1.8,11.2){\line(1,0){1}} \put(-1.8,11.2){\line(0,1){1}}
\put(-0.8,12.2){\line(-1,0){1}} \put(-0.8,12.2){\line(0,-1){1}}

\put(-1.57,11.56){VI$_0$}


\put(4,10.25){\line(-3,-1){2}}  \put(4,10.25){\line(3,-1){2}}
\put(4,8.9){\line(3,1){2}} \put(4,8.9){\line(-3,1){2}}
\put(2.7,9.45){$ \nu_2 >0, \  \nu_3 =0$}

\put(4,10.9){\vector(0,-1){0.65}}

\put(4.2,10.6){no}


\put(0.1,10){$a\!=\!0$} \put(1.25,9.7){yes}

\put(0.5,10.1){\oval(1.1,0.5)} \put(0.5,9.84){\line(0,-1){0.25}}


\put(7.2,10){$a\!\neq\!0$} \put(6.25,9.7){yes}

\put(7.6,10.1){\oval(1.1,0.5)} \put(7.6,9.84){\line(0,-1){0.25}}


\put(6,9.58){\vector(1,0){2.78}} \put(2,9.58){\vector(-1,0){2.78}}

\put(8.8,9.2){\line(1,0){1}} \put(8.8,9.2){\line(0,1){1}}
\put(9.8,10.2){\line(-1,0){1}} \put(9.8,10.2){\line(0,-1){1}}

\put(8.92,9.56){VII$_h$}

\put(-1.8,9.2){\line(1,0){1}} \put(-1.8,9.2){\line(0,1){1}}
\put(-0.8,10.2){\line(-1,0){1}} \put(-0.8,10.2){\line(0,-1){1}}

\put(-1.65,9.56){VII$_0$}
\put(4,8.9){\vector(0,-1){0.65}}

\put(4.2,8.6){no}



\put(4,8.25){\line(-3,-1){2}}  \put(4,8.25){\line(3,-1){2}}
\put(4,6.9){\line(3,1){2}} \put(4,6.9){\line(-3,1){2}}
\put(2.7,7.45){$ \nu \, \nu_3 >0, \  \nu_2>0$}

\put(4,6.9){\vector(0,-1){0.7}}

\put(4.2,6.6){no}


 \put(1.25,7.7){yes}



\put(2,7.58){\vector(-1,0){2.78}}

\put(-1.8,7.2){\line(1,0){1}} \put(-1.8,7.2){\line(0,1){1}}
\put(-0.8,8.2){\line(-1,0){1}} \put(-0.8,8.2){\line(0,-1){1}}

\put(-1.5,7.56){IX}

\put(-1.8,5.7){\line(1,0){1}} \put(-1.8,5.7){\line(0,1){1}}
\put(-0.8,6.7){\line(-1,0){1}} \put(-0.8,6.7){\line(0,-1){1}}

\put(-1.67,6.06){VIII}

\put(4,6.2){\vector(-1,0){4.78}}

\end{picture} }
\end{center}

\vspace{-6.1cm}
\caption{This flow diagram allows us to distinguish the Bianchi-Behr types. The initial data are the scalars $\nu$, $\nu_2$ and $\nu_3$ and the vector $a$ defined in (\ref{eseya}, \ref{N-scalars}, \ref{nus}) in terms of the structure tensor $Z$ (or in terms of the matrix $n_{ab}$ and $A_b$).}
\label{figure-1}
\end{figure}


\section{Multiply-transitive actions: $G_6$ and $G_4$}
\label{sec-multiply}

An old result by Bianchi \cite{bianchi} and Fubini \cite{fubini} forbids the maximal isometric action of a $G_5$ on three-dimensional orbits. Thus, only the multiply-transitive actions of a $G_6$ and a $G_4$ are possible. It is known that the maximal dimension of the isometry group occurs in spaces with constant curvature. Thus, for three-dimensional spaces we get a $G_6$ when the Ricci tensor $R$ admits a triple eigenvalue. Moreover, the sign of this eigenvalue distinguishes three different groups:
\begin{theorem} 
\label{theo-G6}
The necessary and sufficient conditions for a three-dimensional Riemannian
space to admit a G$_6$ is that it has constant curvature, $3\,R =
r \, g$, $r = \tr R$. Moreover
\begin{itemize}
\item[(i)] The group G$_6$ is locally diffeomorph to E(3) when $r=0$.
\item[(ii)] The group G$_6$ is locally diffeomorph to SO(4) when $r >0$.
\item[(iii)]  The group G$_6$ is locally diffeomorph to SO(3,1) when $r < 0$.
\end{itemize}
\end{theorem}

On the other hand, Bona and Coll \cite{bonacoll1, bonacoll2} applied the Eisenhart \cite{eisenhart} theorem on the integrability of Killing vector equations to obtain an invariant characterization of the three-dimensional Riemannian metrics admitting a $G_4$: {\em the metric $g$ admits a $G_4$ as the maximal group of isometries if, and only if, its Ricci tensor is algebraically special, with two constant different eigenvalues,  $\alpha \neq \beta$, $\dif \alpha =0$, $\dif \beta =0$, and the eigenvector $u$ associated with the simple eigenvalue is shear-free. Moreover the group $G_4$ is locally diffeomorph to SO(2,1) $\times$ U(1) when $\alpha + \beta < 0$, to $SO(3) \times U(1)$ when $\alpha + \beta >0$, and to III$_{q=0}$ when $\alpha + \beta =0$}.

In the third case, III$_{q=0}$ is Petrov's notation for the G$_4$ groups \cite{petrov}. The second and the first cases correspond, respectively, to a G$_4$ of type VIII and to a particular representative of a G$_4$ of type I in Petrov's notation.

The above statement certainly offers an invariant characterization, but it is not an IDEAL one. We can obtain an explicit and algorithmic labeling if we write the invariant conditions in terms of explicit concomitants of the Ricci tensor. The algebraically special condition and the sign of $\alpha + \beta$ can be expressed in terms of trace of the Ricci tensor and the traces of the tensorial powers of the traceless part of the Ricci tensor. If these invariant scalars are constant, then the Ricci eigenvalues are constant. Finally, taking into account expression (\ref{sigmaOmega}), the shear-free condition for the eigenvector can be stated in terms of a tensorial first-order concomitant of the Ricci tensor. Indeed, a straightforward tensorial calculation leads to:
\begin{theorem} 
\label{theo-G4}
Let $g$ be a three-dimensional Riemannian metric and $R$ its Ricci tensor. Let us define the following Ricci concomitants:
\begin{eqnarray} \label{rNbc}
r \equiv \tr R \, , \quad S \equiv R - \frac13 r g \, , \quad s \equiv \tr S^2 \, , \quad t \equiv \tr S^3 \, . \\ \hspace{-20mm} \label{Sigma}
\Sigma \equiv D - \frac12 (\tr D) \, h \, , \quad D_{ij} \equiv (\nabla R \cdot R)^{k l m} \eta_{m l (i}\, h_{j)k}   \, , \quad h \equiv \frac{1}{t} \Big(2 \, S^3 - \frac43 s \, S \Big) \, .
\end{eqnarray}
The necessary and sufficient conditions for $g$ to admit a G$_4$ as maximal isometry group are: 
\be
6 t^2 = s^3 \not= 0 \, , \qquad \dif r = \dif s =0 \, , \qquad  \Sigma = 0 \, .
\ee
Moreover
\begin{itemize}
\item[(i)]  The group G$_4$ is locally diffeomorph to III$_{q=0}$ when $16 r^3 + 9 t = 0$.
\item[(ii)] The group G$_4$ is locally diffeomorph to SO(2,1) $\times$ U(1) when $16 r^3 + 9 t < 0$.
\item[(iii)] The group G$_4$ is locally diffeomorph to SO(3) $\times$ U(1) when $16 r^3 + 9 t >0$.
\end{itemize}
\end{theorem}


\section{Simply-transitive actions: $G_3$}
\label{sec-simply}

Now we deal with the simply-transitive case, that is, a G$_3$ as maximal group of isometries acting transitively exists. Again, a result by Bona and Coll \cite{bonacoll1, bonacoll2}, which applies the Eisenhart \cite{eisenhart} theorem, offers an invariant characterization: {\em the metric $g$ admits a $G_3$ as maximal group of isometries if, and only if, there exists an orthonormal basis defined by its Ricci tensor, and all the Ricci eigenvalues and all the connection coefficients of this basis are constant.} Now we look for an IDEAL formulation of this statement.

It is worth remarking that in the above characterization we can remove the condition on the Ricci eigenvalues. Indeed, if all the connection coefficients are constant, then the Ricci components in the orthonormal basis are also constant, and thus the Ricci eigenvalues are too.

The orthonormal basis of the above statement exists in two different cases \cite{bonacoll2}: (i) when the Ricci tensor is algebraically general, and (ii) when it is algebraically special with two different eigenvalues and the eigenvector associated with the simple eigenvalue is not shear-free. In any case, a traceless symmetric algebraically general 2-tensor $T$ exists that is invariant by the three Killing vectors $\{\xi_a\}$,  ${\cal L}_{\xi_a} T =0$, for $i=1,2,3$. In the case (i) $T$ is the traceless part of the Ricci tensor, $T=S$; in the case (ii) $T = \Sigma$, where $\Sigma$ is the first order Ricci concomitant defined in (\ref{Sigma}). 

If we use (\ref{siete}) to replace $\nabla \xi_a$ in the invariant conditions ${\cal L}_{\xi_a} T =0$, we obtain 
\be
(\xi_a)^{k} \Big[ \nabla_{k} T_{ij}  +
T^m_{\ j} {Z_{k}}^{n} \eta_{n i m} +
T_{i}^{\ m} {Z_{k}}^{n} \eta_{n j m}
\Big] =0 \, , \quad a = 1,2,3 \, ,
\ee
and consequently,
\begin{equation} \label{nablaT=B}
\nabla_{k} T_{ij} = Z_k^{\ m}  B_{ijm}  \, ,   \qquad B_{ijk} \equiv \eta_{ikn} T^n_{\ j} +  \eta_{jkn} T_i^{\ n} \, .
\end{equation}
Then, a straightforward computation gives
\begin{equation} \label{BB}
B_{kli} B^{kl}_{\ \ j} = 2 H_{ij} \, , \qquad H \equiv - 3 \, T^2 + 2 b \, g \, , \quad b \equiv \tr T^2 \, , 
\end{equation}
and from (\ref{nablaT=B}) and (\ref{BB}) we obtain
\begin{equation} \label{KZH}
K = Z \cdot H \, , \qquad K_{i}^{\ j} \equiv  (\nabla  T \cdot T)_{i k l} \eta^{l k j}   \, .
\end{equation}
Then, we can obtain the structure tensor $Z$ in terms of $T$ and its covariant derivative provided that $H$ is a regular 2-tensor. From the characteristic equation of a traceless symmetric tensor $T$,
\begin{equation} \label{ce}
T^3 - \frac12 b \, T- \frac13 c \, g= 0 \, , \qquad c \equiv \tr T^3 \, ,
\end{equation}
and from expression (\ref{BB}) for $H$, we obtain:
\begin{equation} \label{H_1}
H \cdot [3 b \, T^2 + 6 c \, T + \frac12 b^2 \, g] = e \, g \, ,
\end{equation}
where the scalar $e \equiv   b^3 - 6 c^2$ does not vanish when $T$ is algebraically general. Then, from (\ref{KZH}) we obtain the structure tensor and, consequently, we can state:

\begin{lemma} \label{lemma-Z(T)}
If $T$ is an algebraically general traceless symmetric 2-tensor that is invariant by a transitive group G$_3$, ${\cal L}_{\xi_a} T =0$, then the structure tensor $Z$ of the group can be obtained as:
\begin{equation} \label{Z(T)}
Z= Z(T) \equiv \frac{1}{e} K \! \cdot \! \Big[3 b \, T^2 + 6 c \, T + \frac12 b^2 \, g \Big]  \, . 
 \end{equation}
where
\begin{equation} \label{Z(T)-b}
K_{i}^{\ j} \equiv  (\nabla  T \cdot T)_{i k l} \eta^{l k j}  , \quad b \equiv \tr T^2 \, , \quad c \equiv \tr T^3 \,  , \quad e \equiv   b^3 - 6 c^2  \, .
\end{equation}
Moreover, if $\{u_a\}$ is an oriented orthonormal eigenframe of $T$, then 
\be \label{Z(u)}
Z = \frac12 \epsilon^{a b c } \nabla u_a \cdot u_b \otimes u_c \, .
\ee
\end{lemma}
The expression (\ref{Z(u)}) for the structure tensor $Z$ follows from the direct substitution of $T = \alpha^a u_a \otimes u_a$ in the expression (\ref{Z(T)}) of $Z$.

On the other hand, if we put $\nabla u_a = \omega^b_a \otimes u_b$ in (\ref{Z(u)}) we obtain $Z$ in terms of the connection one-forms $\omega^b_a$, $Z = \sum_{(abc)} \omega_a^b \otimes u_c$. Then, substituting this expression in equation (\ref{nablaZ}) we find that this equation holds if, and only if, all the connection coefficients, $\gamma_{ab}^c = (u_a, \omega^c_b)$,  are constant. We have then: 

\begin{lemma}
All the conection coefficients of an orthonormal basis  $\{u_a\}$ are constant if, and only if, the 2-tensor $Z$ given in (\ref{Z(u)}) fulfills equation (\ref{nablaZ}), that is,
\be \label{DZ}
{\cal C}(Z) = 0 \, , \quad  {\cal C}(Z)_{k i j} \equiv \nabla_{k} Z_{ij} - Z_k^{\ m} (\eta_{i m n} Z^n_{\ j} +  \eta_{j m n} Z_{i}^{\ n} )  \,  .
\ee
\end{lemma}

From the two lemmas above and the result in \cite{bonacoll2} stated in the first paragraph of this section, we obtain:

\begin{theorem} \label{theo-G3}
Let $g$ be a three-dimensional Riemannian metric and let us consider the Ricci concomitants $S$, $s$, $t$ and $\Sigma$ defined in (\ref{rNbc}) and (\ref{Sigma}).
The metric $g$ admits a G$_3$ as maximal group of isometries if, and only if, its Ricci tensor satisfies one the two following conditions:
\begin{itemize}
\item[(i)] $6 t^2 \not= s^3$, and $Z= Z(S)$ given in (\ref{Z(T)}) and (\ref{Z(T)-b}) fulfills equation (\ref{DZ}).
\item[(ii)] $6 t^2 = s^3$, $\Sigma \not = 0$, and $Z= Z(\Sigma)$ given in (\ref{Z(T)}) and (\ref{Z(T)-b}) fulfills equation (\ref{DZ}).
\end{itemize}
\end{theorem}

Note that from the explicit and invariant expression of the structure tensor $Z$ ($Z(S)$ or $Z(\Sigma)$) we can determine explicit and invariant expressions for the tensor $N$ and the vector $a$. And then, we can apply the algorithm presented in figure \ref{figure-1} in order to determine the specific Bianchi-Behr type of the G$_3$. It is worth remarking that not all the types may occur. We will come back to this fact in next section. 

\begin{figure}[b]

\vspace*{0mm}

\hspace*{8mm} \setlength{\unitlength}{0.9cm} {\small \noindent
\begin{picture}(0,18)
\thicklines

\put(4.5,17){\line(-4,-1){1.5}}
 \put(1.5,17){\line(4,-1){1.5}}
\put(1.5,17){\line(0,1){1}} \put(4.5,18){\line(-1,0){3}}
\put(4.5,18){\line(0,-1){1}} \put(1.6,17.5){$ \ \ g , \ \R  , \ S ,
\ \Sigma,  $}
 \put(2.2,17 ){$ \ \r , \ s  , \ t   $}
\put(3,16.65){\vector(0,-1){0.4}}

\put(3,16.25){\line(-2,-1){1.25}} \put(3,16.25){\line(2,-1){1.25}}
\put(3,15){\line(2,1){1.25}} \put(3,15){\line(-2,1){1.25}}
\put(2.3,15.5){$3 \R   = \r g $}


\put(3,14.25){\line(-2,-1){1.25}} \put(3,14.25){\line(2,-1){1.25}}
\put(3,13){\line(2,1){1.25}} \put(3,13){\line(-2,1){1.25}}
\put(2.25,13.45){$6 t^2 = s^3 $}

\put(3,15){\vector(0,-1){0.75}}


\put(6.75,14.25){\line(-2,-1){1.25}}
\put(6.75,14.25){\line(2,-1){1.25}} \put(6.75,13){\line(2,1){1.25}}
\put(6.75,13){\line(-2,1){1.25}} \put(6.22,13.5){$\Sigma = 0 $}


\put(11.5,14.4){\line(-2,-1){1.5}} \put(11.5,14.4){\line(2,-1){1.5}}
\put(11.5,12.9){\line(2,1){1.5}} \put(11.5,12.9){\line(-2,1){1.5}}
\put(10.5,13.5){$\dif r \! =\! \dif s \! =0$}




\put(4.25,15.63){\vector(1,0){10.25}}
\put(4.25,13.630){\vector(1,0){1.25}} \put(8,13.63){\vector(1,0){2}}
\put(13,13.65){\vector(1,0){1.5}}


\put(14.5,15.2){\line(1,0){1}} \put(14.5,15.2 ){\line(0,1){1}}
\put(15.5,16.2 ){\line(-1,0){1}} \put(15.5,16.2 ){\line(0,-1){1}}
\put(14.8,15.6){G$_6$}


\put(14.5,13.2){\line(1,0){1}} \put(14.5,13.2 ){\line(0,1){1}}
\put(15.5,14.2 ){\line(-1,0){1}} \put(15.5,14.2 ){\line(0,-1){1}}
\put(14.8,13.6){G$_4$}


\put(6.75,13){\vector(0,-1){0.5}} \put(6.75,12.5){\line(0,-1){0.5}}
 \put(3,13){\vector(0,-1){0.5}}
\put(11.5,12.9){\vector(0,-1){0.5}} \put(3,12.5){\line(0,-1){0.5}}
\put(4.87,9.5){\vector(0,-1){0.5}}


\put(-0.72,11.5){\line(1,0){2.75}} \put(-0.73,11.5){\line(0,1){1}}
\put(-0.72,12.5){\line(1,0){2.75}}

\put(2,12.5){\line(1,-1){0.5}}
 \put(2,11.5){\line(1,1){0.5}}
 \put(-0.65,11.85){$Z\! \equiv\! Z(S), \, {\cal C}(Z)$}



\put(7.75,11.5){\line(1,0){2.77}} \put(10.5,11.5){\line(0,1){1}}
\put(7.75,12.5){\line(1,0){2.77}}


 \put(7.52,11.85){$Z\! \equiv \! Z(\Sigma), \, {\cal C}(Z) $}
\put(7.75,12.5){\line(-1,-1){0.5}}
 \put(7.75,11.5){\line(-1,1){0.5}}


\put(4.87,11){\line(-2,-1){1.5}} \put(4.87,11){\line(2,-1){1.5}}
\put(4.87,9.5){\line(2,1){1.5}} \put(4.87,9.5){\line(-2,1){1.5}}
\put(4.,10.1){${\cal C}(Z) =0$}


 \put(3,12){\line(4,-1){1.9}}
\put(6.75,12){\line(-4,-1){1.9}}
\put(4.87,11.55){\vector(0,-1){0.55}}



\put(6.35,10.25){\vector(1,0){8.15}}


\put(14.5,9.7){\line(1,0){1}} \put(14.5,9.7 ){\line(0,1){1}}
\put(15.5,10.7 ){\line(-1,0){1}} \put(15.5,10.7 ){\line(0,-1){1}}
\put(14.8,10.1){G$_3$}

\put(2.5,12){\line(1,0){0.5}}

\put(6.75,12){\line(1,0){0.5}}

\put(11.4,11.8){{\large $\nexists$}}
 \put(4.7,8.4){{\large $\nexists$}}

\put(4.55,15.8){yes}
\put(8.7,13.8){yes}
 \put(6.7,10.45){yes}
\put(4.45,13.8){yes} \put(13.45,13.8){yes}

\put(3.2,12.5){no}
 \put(6.16,12.5){no}
 \put(3.2,14.5){no}
\put(11.65,12.55){no}

\put(5.05,9.15){no}

\end{picture} }  

\vspace{-77mm}
\caption{This flow diagram distinguishes the dimension of the groups that act transitively. We use as initial input data the metric $g$, the Ricci tensor $R$, the algebraic Ricci concomitants $S$, $r$, $s$ and $t$ defined in (\ref{rNbc}), and the first-order Ricci concomitant $\Sigma$ defined in (\ref{Sigma}). In the last step we need two new Ricci concomitants: (i) if $6 t^2 \not= s^3$, $Z=Z(S)$ (first-order) and ${\cal C}(Z)$ (second-order); (ii) if $6 t^2 = s^3$, $Z=Z(\Sigma)$ (second-order) and ${\cal C}(Z)$ (third-order). The diagram has three horizontal end arrows that fall into a G$_6$, a G$_4$ or a G$_3$ as the maximal transitive group. The two vertical end arrows lead to non-existence of a transitive group.}
\label{figure-2}
\end{figure}
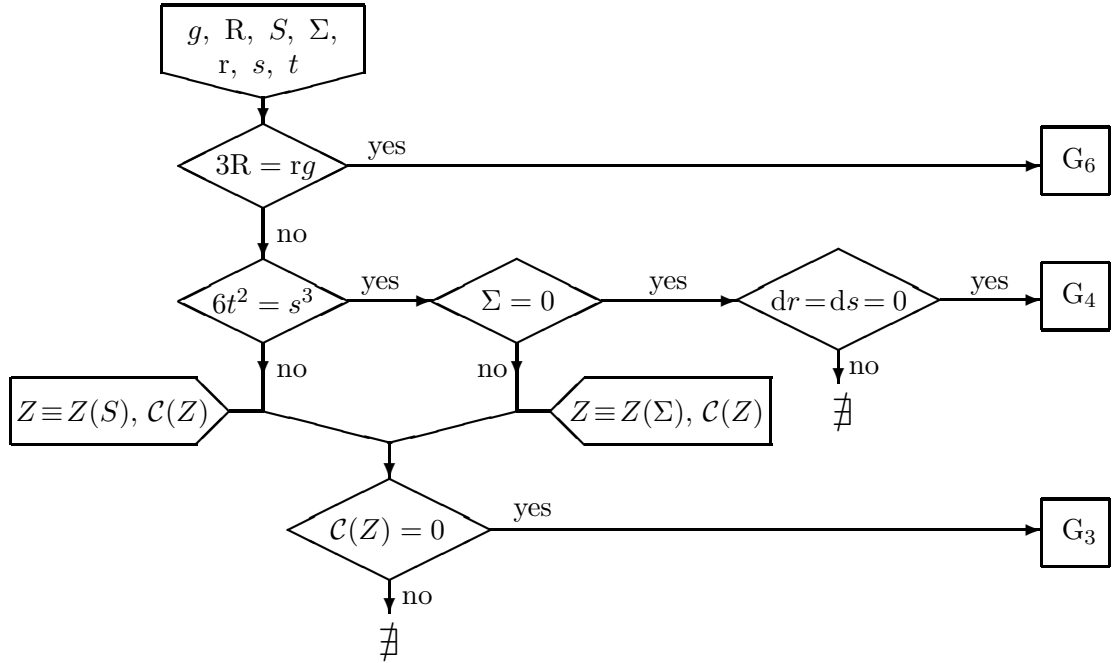

Now, with the results of the last two sections we can built a flow chart that anables us to determine whether an isometry group is acting transitively on a three-dimensional Riemannian space, and to distinguish their dimension (see figure \ref{figure-2}).


\section{Ideal labeling of the Bianchi-Behr types}
\label{sec-ideal-BB}

It is known that several Bianchi types cannot maximally act in a three-dimensional Riemannian space \cite{bianchi} (see also \cite{bonacoll2}). Thus, Bianchi types I and V imply orbits of constant curvature and, consequently, the maximal isometry group is a G$_6$. And Bianchi types II and III imply a G$_4$ as maximal isometry group. Now we recover these results and we analyze the compatibility of each Bianchi-Behr type with the dimension of the maximal group of isometries. Our study confirms and improves the results by Ellis and MacCallum \cite{Ellis-McCallum} and leads to an IDEAL labeling of each Bianchi-Behr type.


\subsection{Metrics admitting a G$_6$}

The Riemannian metrics admitting a G$_6$ as isometry group are characterized by a unique real parameter: its constant curvature $\kappa=3 r$. And the sign of this curvature characterizes the three possible groups as stated in theorem \ref{theo-G6}. In \ref{A-Ricci} we have shown (proposition \ref{propo-ricci1}) that if a Bianchi type I (respectively, type V) acts transitively, then the Ricci tensor vanishes (respectively, is $3 R =r g$, $r < 0$). 

On the other hand, proposition \ref{propo-ricci-7} states that if a Bianchi type VII acts transitively and $N$ has two equal eigenvalues, then the Ricci tensor has three equal eigenvalues ($3 R =r g$). Moreover $r = 0$ for type VII$_{0}$, and $r < 0$ for type VII$_h$.

Finally, proposition \ref{propo-ricci-8-9} states that if a Bianchi type IX acts transitively and $N$ has three equal eigenvalues, then the Ricci tensor has also three equal positive eigenvalues ($3 R =r g$, $r > 0$).
Consequently, we obtain:
\begin{theorem} \label{theo-G6-BB}
(i) A three-dimensional Riemannian space admits a group G$_6$ with $r =0$ (diffeomorf to E(3)) if, and only if, it admits a transitive G$_3$ of Bianchi-Behr type I. Moreover, then it admits a transitive group G$_3$ of Bianchi-Behr type VII$_{0}$. \\
(ii) A three-dimensional Riemannian space admits a group G$_6$ with $r < 0$ (diffeomorf to SO(3,1)) if, and only if, it admits a transitive G$_3$ of Bianchi-Behr type V. Moreover, then it admits a transitive group G$_3$ of Bianchi-Behr type VII$_{h}$. \\
(iii) If a three-dimensional Riemannian space admits a group G$_6$ with $r > 0$ (diffeomorf to SO(4)) then it admits a transitive group G$_3$ of Bianchi-Behr type IX.  
\end{theorem}
Note that Bianchi-Behr types I and V can only act transitively when the metric admits a G$_6$ as maximal group of isometries. Nevertheless, Bianchi-Behr types VII$_0$, VII$_h$ and IX act transitively when a G$_6$ exists, but they can also act transitively as maximal group, and type IX can acts transitively when the maximal group is a G$_4$ (see following subsections). The results of the above theorem can be given in the algorithmic form presented in figure \ref{figure-3}.


\begin{figure}[t]

\vspace*{0mm}
\hspace*{45mm} \setlength{\unitlength}{0.9cm} {\small \noindent
\begin{picture}(0,18)
\thicklines


\put(-1,14.625){\line(1,0){1}} \put(-1,14.625){\line(0,1){0.75}}
\put(0,15.375 ){\line(-1,0){1}} \put(0,15.375 ){\line(0,-1){0.75}}
\put(-0.8,14.825){{\large{G$_6$}}}


\put(7,17.25){\line(1,0){1}} \put(7,17.25){\line(0,1){0.75}}
\put(8,18  ){\line(-1,0){1}} \put(8,18  ){\line(0,-1){0.75}}
\put(7.3,17.5){BI}

\put(7.5,16.5){\oval(1.5,0.75)} \put(7,16.35 ){BVII$_0$}


\put(7,15){\line(1,0){1}} \put(7,15){\line(0,1){0.75}} \put(8,15.75
){\line(-1,0){1}} \put(8,15.75 ){\line(0,-1){0.75}}
\put(7.22,15.23){BV}

\put(7.5,14.25){\oval(1.5,0.75)} \put(6.97,14.1 ){BVII$_h$}


\put(7.5,13.15){\oval(1.5,0.75)} \put(7.14,13 ){BIX}


\put(3,14.6){\line(1,0){1.75}} \put(3,14.6){\line(0,1){0.75}}
\put(4.75,15.35 ){\line(-1,0){1.75}} \put(4.75,15.35
){\line(0,-1){0.75}} \put(3.2,14.85){SO(3,1)}


\put(3,12.75){\line(1,0){1.75}} \put(3,12.75){\line(0,1){0.75}}
\put(4.75,13.5 ){\line(-1,0){1.75}} \put(4.75,13.5
){\line(0,-1){0.75}} \put(3.4,13){SO(4)}


\put(3,16.5){\line(1,0){1.75}} \put(3,16.5){\line(0,1){0.75}}
\put(4.75,17.25 ){\line(-1,0){1.75}} \put(4.75,17.25
){\line(0,-1){0.75}} \put(3.55,16.75){E(3)}


\put(4.75,16.9  ){\line(1,0){0.75}} \put(5.5,16.9){\line(0,1){0.75}}
\put(5.5,16.9 ){\line(0,-1){0.4}}
 \put(5.49,17.65){\line(1,0){1.5}}
\put(5.49,16.5){\line(1,0){1.25}}

 \put(5.49,17.55){\line(1,0){1.5}}


\put(4.75,14.9  ){\line(1,0){0.75}} \put(5.5,14.9){\line(0,1){0.55}}
\put(5.5,14.9 ){\line(0,-1){0.55}}
 \put(5.49,15.45){\line(1,0){1.5}}
\put(5.49,14.35){\line(1,0){1.25}}

 \put(5.49,15.35){\line(1,0){1.5}}

\put(4.75,13.15  ){\line(1,0){2 }}


\put(1.4,15.1){$r\!<\!0$} \put(1.4,17){$r\!=\!0$}
\put(1.4,13.2){$r\!>\!0$}


\put(1,15 ){\line(1,0){2}} \put(1,15 ){\line(0,1){1.9}}\put(1,15
){\line(0,-1){1.9}} \put(1,13.1){\line(1,0){2}}
\put(1,16.9){\line(1,0){2}}

\put(0,15 ){\line(1,0){2}}
\end{picture} }
\vspace*{-11.5cm}
\caption{Once we know that a G$_6$ exists (see flow chart in figure \ref{figure-2}), the sign of the curvature determines the specific isometry group. This diagram also shows the Bianchi-Behr subgroups (that act transitively) admitted in every case. The double line indicates the equivalence between the existence of a specific G$_6$ and the existence of a specific Bianchi-Behr G$_3$ group.}
\label{figure-3}
\end{figure}



\subsection{Metrics admitting a G$_4$}

The Riemannian metrics admitting a G$_4$ as isometry group have an algebraically special Ricci tensor and they are characterized by two real parameters: the two constant Ricci eigenvalues $\alpha$ and $\beta$ (or, equivalently, the scalar invariants $r$ and $t$ defined in theorem \ref{theo-G4}). This theorem characterizes the maximal action of a G$_4$ and the sign of the invariant $\alpha + \beta$ (or, equivalently, the sign of $16 r^3 + 9t$) distinguishes the three possible groups. Note that, these two parameters are not arbitrary. Indeed, as we have shown in \ref{A-AS} (see proposition \ref{propo-sigma0})  the simple eigenvalue is, necessarily, non negative, $\alpha \geq 0$. Moreover, $\alpha >0$ when $\alpha + \beta =0$. The explicit expression of the simple eigenvalues in terms of the scalar invariants given in (\ref{rNbc}) is $\alpha \equiv 2 t/s + r/3$.
%
%

In \ref{A-Ricci} we have shown (propositions \ref{propo-ricci-2} and \ref{propo-ricci-3}) that if a Bianchi type II (respectively, type III) acts transitively, then the Ricci tensor is algebraically special and $\alpha+\beta=0$, and $\alpha >0$ (respectively, $\alpha+\beta<0$, and $\alpha \geq 0$). Moreover, proposition \ref{propo-sigma-2-3} in \ref{A-AS} states that in both cases the simple eigenvector is shear-free.

On the other hand, proposition \ref{propo-ricci-8-9} states that if a Bianchi type VIII (respectively, type IX) acts transitively and $N$ has two equal eigenvalues, then the Ricci tensor is algebraically special with $\alpha > 0$ and $\alpha+\beta<0$ (respectively, $\alpha+\beta>0$). Moreover, proposition \ref{propo-sigma-6-8-9} states that in both cases the simple eigenvector is shear-free. Consequently, we obtain:
\begin{theorem}  \label{theo-G4-BB}
(i) A three-dimensional Riemannian space admits a maximal group G$_4$ with $16 r^3 + 9 t = 0$ (diffeomorf to III$_{q=0}$) if, and only if, it admits a transitive G$_3$ of Bianchi-Behr type II. \\
(ii) A three-dimensional Riemannian space admits a maximal group G$_4$ with $16 r^3 + 9 t < 0$ (diffeomorf to SO(2,1) $\times$ U(1)) if, and only if, it admits a transitive G$_3$ of Bianchi-Behr type III. Moreover, when $\alpha \equiv 2 t/s + r/3  >0$ it admits a transitive group G$_3$ of Bianchi-Behr type VIII. \\
(iii) If a three-dimensional Riemannian space admits a maximal group G$_4$ with $16 r^3 + 9 t > 0$ (diffeomorf to SO(3) $\times$ U(1)) then: when $\alpha >0$ it admits a transitive group G$_3$ of Bianchi-Behr type IX; when $\alpha = 0$ it does not admit any G$_3$ in transitive action.
\end{theorem}
Note that Bianchi-Behr types II and III can only act transitively when the metric admits a G$_4$ as the maximal group of isometries. Nevertheless, Bianchi-Behr types VIII and IX can act transitively when a G$_4$ exists, but they can also act transitively as maximal group (see following subsection), and type IX can act transitively when the maximal group is a G$_6$ as we have seen in the section above. The results of the above theorem can be given in the algorithmic form presented in figure \ref{figure-4}.


\begin{figure}

\vspace*{-3mm}

\hspace*{40mm} \setlength{\unitlength}{0.9cm} {\small \noindent
\begin{picture}(0,18)
\thicklines


\put(-1,14.625){\line(1,0){1}} \put(-1,14.625){\line(0,1){0.75}}
\put(0,15.375 ){\line(-1,0){1}} \put(0,15.375 ){\line(0,-1){0.75}}
\put(-0.8,14.825){{\large{G$_4$}}}

\put(1.4,15.17){$\gamma \!<\!0$} \put(1.4,17.07){$\gamma  \!=\!0$}
\put(1.4,13.27){$\gamma \!>\!0$}


\put(3,14.6){\line(1,0){3}} \put(3,14.6){\line(0,1){0.75}}
\put(6,15.35 ){\line(-1,0){3}} \put(6,15.35 ){\line(0,-1){0.75}}
\put(3.3,14.85){SO(2,1)$\!\times\!$U(1)}

\put(3,12.75){\line(1,0){3}} \put(3,12.75){\line(0,1){0.75}}
\put(6,13.5 ){\line(-1,0){3}} \put(6,13.5 ){\line(0,-1){0.75}}
\put(3.4,13){SO(3)$\!\times\!$U(1)}


\put(3,16.5){\line(1,0){1.75}} \put(3,16.5){\line(0,1){0.75}}
\put(4.75,17.25 ){\line(-1,0){1.75}} \put(4.75,17.25
){\line(0,-1){0.75}} \put(3.35,16.75){III$_{q=0}$}



\put(8.25,15.25){\line(1,0){1}} \put(8.25,15.25){\line(0,1){0.75}}
\put(9.25,16 ){\line(-1,0){1}} \put(9.25,16 ){\line(0,-1){0.75}}
\put(8.36,15.5){BIII}

\put(8.75,14.35){\oval(1.5,0.75)} \put(8.22,14.2 ){BVIII}


\put(6,14.9  ){\line(1,0){0.75}} \put(6.75,14.9){\line(0,1){0.75}}
\put(6.75,14.9 ){\line(0,-1){0.55}}
\put(6.75,15.65){\line(1,0){1.5}} \put(6.75,14.35){\line(1,0){1.25}}

 \put(6.75,15.55){\line(1,0){1.5}}

\put(8.25,16.5){\line(1,0){1}} \put(8.25,16.5){\line(0,1){0.75}}
\put(9.25,17.25  ){\line(-1,0){1}} \put(9.25,17.25
){\line(0,-1){0.75}} \put(8.5,16.75){BII}




\put(8.75,13.15){\oval(1.5,0.75)} \put(8.4,13 ){BIX}


 \put(6.94 ,13.3 ){$\alpha\!>\!0$}
 \put(6.94 ,14.5 ){$\alpha\!>\!0$}



\put(4.75,16.95){\line(1,0){3.5}} \put(4.75,16.85){\line(1,0){3.5}}

\put(6,13.15  ){\line(1,0){2 }}


\put(1,15 ){\line(1,0){2}} \put(1,15 ){\line(0,1){1.9}} \put(1,15
){\line(0,-1){1.9}} \put(1,13.1){\line(1,0){2}}
\put(1,16.9){\line(1,0){2}}

\put(0,15 ){\line(1,0){2}}
\end{picture} }
\vspace*{-115mm}
\caption{Once we know that a G$_4$ exists (see flow chart in figure \ref{figure-2}), the sign of the scalar $\gamma \equiv 16 r^3 + 9t$ determines the specific isometry group. This diagram also shows the Bianchi-Behr subgroups (that act transitively) admitted in every case. The double line indicates the equivalence between the existence of a specific G$_4$ and the existence of a specific Bianchi-Behr G$_3$ group. The scalar $\alpha \equiv 2 t/s + r/3$ is positive when $\gamma =0$ and non-negative when $\gamma \not=0$. But only when $\alpha >0$ the Bianchi-Behr types VIII and IX act transitively.}
\label{figure-4}
\end{figure}



\subsection{Metrics admitting a G$_3$ with an algebraically special Ricci tensor}

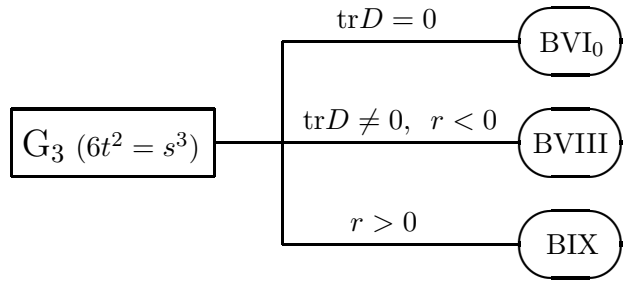
\begin{figure}

\vspace*{-10mm}

\hspace*{40mm} \setlength{\unitlength}{0.9cm} {\small \noindent
\begin{picture}(0,18)
\thicklines


\put(0,14 ){\line(1,0){3}} \put(0,14 ){\line(0,1){1}} \put(3,15
){\line(-1,0){3}} \put(3,15  ){\line(0,-1){1}}
\put(0.15,14.3){{\large{G$_3$}}\ ($6t^2 =s^3$)}


\put(3,14.5 ){\line(1,0){4.5}}

\put(4,14.5 ){\line(0,1){1.5}}

\put(4,14.5 ){\line(0,-1){1.5}}

\put(4,16 ){\line(1,0){3.5}}

\put(4,13 ){\line(1,0){3.5}}

\put(4.30,14.7 ){$\tr \! D \neq 0$, $\ r <0$}

\put(4.8,16.2 ){$\tr \! D = 0$}

\put(5,13.2 ){$r > 0$}

\put(8.25,13 ){\oval(1.5,1)} \put(7.9,12.85 ){BIX}

 \put(8.25,14.5){\oval(1.5,1)} \put(7.7,14.35 ){BVIII}

 \put(8.25,16){\oval(1.5,1)} \put(7.8,15.85 ){BVI$_0$}

\end{picture} }

\vspace*{-115mm}
\caption{Once we know that a G$_3$ with an algebraically special Ricci tensor exists (see flow chart in figure \ref{figure-2}), the sign of the scalar invariants $\tr \! D$ and $r$ determines the specific Bianchi-Behr group that acts transitively as this diagram shows.}
\label{figure-5}
\end{figure}

As we have shown in propositions \ref{propo-ricci-6}, \ref{propo-ricci-8-9} (\ref{A-Ricci}) and propositions \ref{propo-sigmanot0} and \ref{propo-sigma-6-8-9} (\ref{A-AS}) Bianchi-Behr types VI$_0$, and VIII and IX (with the eigenvalues of $N$ constrained by $\varphi_2 + \varphi_3 = \varphi_1 \not= 2 \varphi_2$) act transitively on Riemannian metrics with an algebraically special Ricci tensor, and the simple eigenvector having a non-vanishing shear. This means that the action is maximal. Moreover, the double eigenvalue vanishes, and then the metric depends on two parameters, the simple eigenvalue $\alpha = r = \tr R$, and the rotation scalar $\Omega$ (and then $\alpha = 2(\Omega^2- \sigma^2)$). Furthermore, $\alpha > 0$ for type IX and $\alpha < 0$ for types VI$_0$ and VIII. And $\Omega =0$ only for type VI$_0$. The scalar invariant $\Omega$ can be obtained from the Ricci concomitant $D$ given in (\ref{sigmaOmega}) as $\Omega = - \frac12 (\alpha-\beta)^{-2} \tr D$. Thus, $\Omega=0$ if, and only if, $\tr D =0$. Consequently, we obtain:
\begin{theorem}  \label{theo-G3AS-BB}
If a three-dimensional Riemannian space with an algebraically special Ricci tensor admits a G$_3$ as maximal group of isometries, then:\\
(i) The group is a Bianchi-Behr type VI$_0$ if, and only if, $\tr \! D = 0$. \\
(ii) The group is a Bianchi-Behr type VIII if, and only if, $\tr \! D \not= 0$ and $r<0$. \\
(iii) The group is a Bianchi-Behr type IX if, and only if, $r > 0$. 
\end{theorem}
It is worth remarking that Bianchi-Behr types VI$_0$, VIII and IX can also act maximally when the Ricci tensor is algebraically general (see following subsection). The results of the above theorem can be given in the algorithmic form presented in figure \ref{figure-5}.

\subsection{Metrics admitting a G$_3$ with an algebraically general Ricci tensor}

If the Bianchi-Behr types IV and VI$_h$ act transitively, then the Ricci tensor is algebraically general (see propositions \ref{propo-ricci-4} and \ref{propo-ricci-6} in \ref{A-Ricci}). The other Bianchi-Behr types compatible with the algebraic general condition are types VI$_0$, VII$_h$, VII$_0$, VIII and IX (see propositions \ref{propo-ricci-6}, \ref{propo-ricci-7} and \ref{propo-ricci-8-9}). Now we can obtain the structure tensor $Z$ as the concomitant of the Ricci tensor $Z \equiv Z(S)$ given in (\ref{Z(T)}) and (\ref{Z(T)-b}). Consequently, the different Bianchi-Behr types can be distinguished by using the algorithm presented in figure \ref{figure-1} (subsection \ref{subsec-algo-BB}). Consequently, we obtain:
\begin{theorem}  \label{theo-G3AG-BB}
Let $g$ be a three-dimensional Riemannian metric with an algebraically general Ricci tensor that admits a G$_3$ as maximal group of isometries. Let us consider the Ricci concomitants $\nu$, $\nu_2$, $\nu_3$ and $a$ defined in (\ref{eseya}), (\ref{N-scalars}) and (\ref{nus}) in terms of $Z=Z(S)$ given in (\ref{Z(T)}) and (\ref{Z(T)-b}). Then:\\
(i) The group is a Bianchi-Behr type IV if, and only if, $\nu \not=0, \ \nu_2 = \nu_3=0$. \\
(ii) The group is a Bianchi-Behr type VI$_0$ (respectively, type VI$_h$, $h \not= -1$) if, and only if, $\nu_2 < 0, \ \nu_3=0$, and $a=0$ (respectively, $a \not=0$). \\
(iii) The group is a Bianchi-Behr type VII$_0$ (respectively, type VII$_h$) if, and only if, $\nu_2 > 0, \ \nu_3=0$, and $a=0$ (respectively, $a \not=0$). \\
(iv) The group is a Bianchi-Behr type IX if, and only if, $\nu_2 > 0$ and $\nu \, \nu_3>0$. \\
(v) The group is a Bianchi-Behr type VIII if, and only if, either $\nu_2 \leqslant 0$ or $\nu \, \nu_3 \leqslant 0$. \\
\end{theorem}
Note that in the Bianchi-Behr classification (see table \ref{table-2} and figure \ref{figure-1}) type VI$_h$ includes the Bianchi type III ($h=-1$). Nevertheless, in our study type III implies a G$_4$ (see theorem \ref{theo-G4}) and, consequently, we must exclude the value $h=-1$ in point (ii) of the above theorem.


\section{Comments and work in progress}
\label{sec-comments}

In this paper we have presented an invariant approach to the homogeneous three-dimensional Riemannian spaces. On one hand, we have improved the invariant study by Bona and Coll \cite{bonacoll1, bonacoll2} that presented the necessary and sufficient conditions for the metric to admit a  group G$_r$ on s-dimensional orbits. Here, we have considered the transitive actions (G$_6$, G$_4$ and G$_3$) and we have offered an explicit expression of these conditions and so, we have obtained an IDEAL labeling of these geometries (theorems \ref{theo-G6}, \ref{theo-G4} and \ref{theo-G3}). Then, a flow diagram can be drawn up to detect each case in an algorithmic form (see figure \ref{figure-2} at the end of section \ref{sec-simply}).

On the other hand, we have studied the Bianchi-Behr types that can maximally act and those that are compatible with G$_6$ or G$_4$ as maximal groups of isometries. This analysis leads to a classification of the homogeneous three-dimensional Riemann spaces. Each of the classes is characterized in terms of invariant and explicit conditions that give the full IDEAL labeling of the different Bianchi-Behr types (theorems \ref{theo-G6-BB}, \ref{theo-G4-BB},  \ref{theo-G3AS-BB} and \ref{theo-G3AG-BB}). Again, diagrams can be built to detect each case in an algorithmic form (see figures \ref{figure-1}, \ref{figure-2} and \ref{figure-3}). 

Our study shows that there are some Bianchi-Behr types that necessarily belong to one of the considered classes. And there are others that are compatible with more than one class. In any case, from the results in section \ref{sec-ideal-BB}, a specific algorithmic labeling easily follows for the different Bianchi-Behr types. For example, type I is characterized by a vanishing Ricci tensor $R=0$ (the metric admits the six-dimensional E(3) as isometry group). But type IX is compatible with four classes and requires a more complex algorithm to label it. Note that the unfolding of some Bianchi-Behr types is a consequence of their action on the Riemannian space. Indeed, in the usual Bianchi-Behr classification one considers the rank and signature of the tensor $N$ (or of the matrix $n^{ab}$). But now, the metric tensor allows us to analyze $N$ as an endomorphism and to study its eigenvalues and eigenvectors.

It is worth remarking that a family of homogeneous three-dimensional Riemannian metrics exists that does not admit a Bianchi-Behr group acting transitively. Indeed, as stated in theorem \ref{theo-G4-BB}, if  SO(3) $\times$ U(1)) acts as maximal (four-dimensional) group of isometries, then the Ricci tensor is algebraically special and the simple eigenvalue is non negative, $\alpha \geqslant 0$. The group of Bianchi-Behr type IX is a a subgroup of SO(3) $\times$ U(1) that acts transitively when $\alpha >0$, but it has two-dimensional orbits when $\alpha = 0$. The latter condition implies that the simple eigenvector of the Ricci tensor is covariantly constant. This case was previously mentioned in \cite{Ellis-McCallum}.

Our IDEAL approach to the homogeneous three-dimensional Riemannian spaces is in itself of interest conceptually speaking. Moreover, it is a first necessary step in tackling a similar study for the spatially homogeneous spacetimes. This task is currently in progress and will allow us to obtain an IDEAL labeling of the Bianchi models.

The IDEAL characterization of spacetimes provides
an algorithmic way to test if a metric tensor, given in an arbitrary
coordinate system, is a specific solution of Einstein equations.
The  seminal results by Cartan \cite{cartan} and a deep knowledge of the algebraic structure of the Ricci and Weyl tensors \cite{bcm, fms} both play a central role in achieving the IDEAL characterization for any given metric. By using these results, we have performed an approach that has been useful in labeling, among others, the Schwarzschild \cite{fsS}, Reissner-Nordstr\"om \cite{fsD} and Kerr \cite{fsKerr} black holes, the Lema\^itre-Tolman-Bondi \cite{fs-SSST} or Bertotti-Robinson \cite{fswarped} solutions, and the Stephani or the Szekeres-Szafron universes \cite{fs-SSST-Ricci, fs-Szafron-Ideal}. Other authors have presented IDEAL approaches for higher dimensional spaces \cite{Khavkine, Khavkine-b}. These studies show the interest in obtaining a fully algorithmic characterization of the initial data which correspond to a given solution \cite{garcia-parrado-vk, garcia-parrado, garcia-parrado-2016}.


\ack This work has been partially supported by the Spanish ``Ministerio de
Econom\'{\i}a y Competitividad", MINECO-FEDER project FIS2015-64552-P.


\appendix

\section{Ricci tensor of the Bianchi-Behr types}
\label{A-Ricci}

Here we study the algebraic type of the Ricci tensor of a three-dimensional Riemannian space when a G$_3$ of a specific Bianchi-Behr type acts transitively. We make use of the expression (\ref{ricci2}) that gives the Ricci tensor in terms of $N$ and $a$, and we consider their expression for the different Bianchi types given in table \ref{table-1}. We analyze the different possible ranks of $N$ separately. 

Firstly, we consider the cases with $N=0$, that is, type I if $a=0$, and type V if $a\not=0$. Then, from (\ref{ricci2}) we obtain:

\begin{proposition} \label{propo-ricci1}
(i) If a Bianchi type I acts transitively, then the Ricci tensor vanishes.\\
(ii) If a Bianchi type V acts transitively, then the Ricci tensor is $R = -2 a^2 g$. 
\end{proposition}

Secondly, we consider the cases when $N$ has rank one or two, that is, types II, VI$_0$ and VII$_0$ if $a=0$, and types IV, VI$_h$ (including type III) and VII$_h$ if $a\not=0$. As table \ref{table-1} shows, in all these types an abelian G$_2$ exists, $[\xi_1, \xi_2]=0$, and if $u$ denotes the unitary vector that is orthogonal to its orbits, it holds: 
\begin{equation} \label{ua}
u \propto *(\xi_1 \wedge \xi_2) \, , \qquad N(u) = 0 \, , \qquad  u \wedge a =0 \, .
\end{equation}
Now, $N$ can be considered as an endomorphism in the plane
spanned by $\xi_1$ and $\xi_2$, and then, an orthonormal
eigenframe $\{ v, w \}$ exists. If $\varphi$ and $\psi$ denote the
eigenvalues, we obtain from (\ref{ricci2}):
\begin{lemma} \label{lemma-ricci2}
In Bianchi types where $N$ has rank one or two, $N = \varphi \, v \otimes v + \psi \, w \otimes w$, the Ricci tensor is
\be \label{riccis} 
\hspace{-2cm} R= (\varphi - \psi) \left[ (\varphi \ v \otimes v  - \psi \ w
\otimes w ) - \sqrt{a^2} ( v \, \widetilde{\otimes}\, w) \right] -
\frac{1}{2} \left[ (\varphi - \psi)^2 + 4 a^2 \right]  g \, ,
\ee
and the Ricci eigenvalues are:
\be \label{riccis-eigen} 
\hspace{-15mm} 
\alpha_{1} = - \frac12 [(\varphi-\psi)^2 + 4a^2] , \qquad \alpha_{\pm} = - 2 a^2  \pm \frac{1}{2} (\varphi-\psi) 
 \sqrt{(\varphi+\psi)^2 + 4 a^2}].
\ee
Consequently, the Ricci is algebraically special if, and only if, one of the
following three conditions hold: 
\begin{itemize}
\item[(i)]  $\varphi =  \psi$, and then $R = - 2 a^2 g$.
\item[(ii)] $a=\varphi +  \psi =0$, and then $R = - \frac12 \varphi^2 \, u \otimes u$, \  
\item[(iii)] $a^2 + \varphi  \psi =0$, and then $R = (\alpha - \beta) \omega \otimes \omega + \beta \, g$,  $\alpha + \beta = -4 a^2 \leqslant 0,\  \alpha = \frac12(\varphi + \psi)^2 \geqslant 0$.
\end{itemize}
\end{lemma}
Note that the case (i) leads to a Ricci tensor with three equal eigenvalues. The case (ii) corresponds with two different eigenvalues (the double one is zero), and the simple eigenvector is orthogonal to the plane $\{\xi_1, \xi_2\}$. And in the case (iii) we have two different eigenvalues, and the simple eigenvector lyes in the plane $\{\xi_1, \xi_2\}$.

For Bianchi type II we have (see table \ref{table-1}) $N = |g|^{-1/2} \xi_1 \otimes \xi_1= \varphi \, v \otimes v$ and $a=0$, that is, $\psi =0$, and the Ricci tensor corresponds to the algebraically special case (iii) of the lemma \ref{lemma-ricci2}. More precisely, we obtain:
\begin{proposition}  \label{propo-ricci-2}
If a Bianchi type II acts transitively, then the Ricci tensor is algebraically special:
\be
R = (\alpha - \beta )\, v \otimes v + \beta \, g \, , \qquad \alpha + \beta = 0 \, , \quad \alpha > 0 \, ,
\ee
and the eigenvector associated with the simple eigenvalue has the directions of a Killing vector, $v \propto \xi_1$.
\end{proposition}
Note that in the Bianchi type II it holds: $\tr N = \varphi \not=0$ and $ r = \tr R = - \alpha < 0$.


For Bianchi type III we have (see table \ref{table-1}) $N = -\frac{1}{2 } |g|^{-1/2} \xi_1
\widetilde{\otimes} \xi_2= \varphi \, v \otimes v + \phi \, w \otimes w$ and $a = -\frac{1}{2} |g|^{-1/2} *(\xi_1
\wedge \xi_2) = \sqrt{a^2} \, u$. A straightforward calculation leads to $\varphi \psi + a^2 = 0$. Then, the Ricci tensor corresponds to the algebraically special case (iii) of the lemma \ref{lemma-ricci2}, and the Killing vector $\xi_2$ is a simple Ricci eigenvector. More precisely, we obtain:
\begin{proposition} \label{propo-ricci-3}
If a Bianchi type III acts transitively, then the Ricci tensor is algebraically special:
\be
R = (\alpha - \beta )\, \omega \otimes \omega + \beta \, g \, , \qquad \alpha + \beta < 0 \, , \qquad \alpha \geqslant 0 \, ,
\ee
and the eigenvector associated with the simple eigenvalue has the direction of a Killing vector, $\omega \propto \xi_2$.\end{proposition}
Note that Bianchi type III admits the particular case $\tr N =0$ (iff $R(a)=0$ iff $\alpha =0$).


For Bianchi type IV we have (see table \ref{table-1})  $N = |g|^{-1/2} \xi_1 \otimes \xi_1 = \varphi \, v \otimes v$ and  $a = |g|^{-1/2}  *(\xi_1 \wedge \xi_2) = \sqrt{a^2} \, u\, $. Then, $\psi =0$, $a^2 \not=0$, and none of the three conditions in lemma \ref{lemma-ricci2} holds. Consequently, the Ricci tensor is algebraically general, and we obtain:
\begin{proposition}  \label{propo-ricci-4}
If a Bianchi type IV acts transitively, then the Ricci tensor is algebraically general.
\end{proposition}
%


For Bianchi type VI we have (see table \ref{table-1}) $a = \frac{1}{2}(q+1) |g|^{-1/2} *(\xi_1 
\wedge \xi_2) = \sqrt{a^2} \, u$ and $N = \frac{1}{2}(q-1) |g|^{-1/2}\ \xi_1  
\widetilde{\otimes} \xi_2= \varphi \, v \otimes v + \phi \, w \otimes w$, $q \not= 0, 1$. A straightforward calculation leads to:
\be
\hspace{-10mm} 
\varphi + \psi = \frac{q-1}{2 \sqrt{|g|}}\, \xi_1 \cdot \xi_2 \, , \qquad \varphi \, \psi = - \frac{(q-1)2}{4 |g|}[\xi_1^{2}\,  \xi_2^{2} - (\xi_1 \cdot \xi_2)^2] < 0  ,
\ee
and $\varphi \not= \psi$, $a^2 + \varphi \, \psi \not=0$. Consequently, lemma \ref{lemma-ricci2} implies that for Bianchi-Behr type VI$_h$, $h \not= -1$, ($a \not=0$) the Ricci tensor is algebraically general. And for Bianchi-Behr type VI$_0$ ($a =0$) the Ricci tensor is algebraically special if, and only if, $\varphi + \psi = 0$, that is, $\xi_1 \cdot \xi_2 = 0$. 
Thus, taking into account the expressions in lemma \ref{lemma-ricci2}, we obtain:
\begin{proposition}  \label{propo-ricci-6}
(i) If a Bianchi-Behr type VI$_h$, $h \not= -1$, acts transitively, then the Ricci tensor is algebraically general.\\
(ii) If a Bianchi-Behr type VI$_0$ acts transitively, then the Ricci tensor is algebraically general if, and only if, $\tr N \not=0$. Otherwise, when $\tr N =0$, the Ricci tensor is:
\be
R = \alpha \, u \otimes u  \, , \qquad  \alpha < 0 \, .
\ee
\end{proposition}
Note that in the Bianchi-Behr type VI$_0$, if the Ricci tensor has an eigenplane, the associated double eigenvalue vanishes and there are two orthogonal Killing vectors lying on it.


For Bianchi type VII we have (see table \ref{table-1})  $a = -\frac{q}{2} |g|^{-1/2} *(\xi_1 
\wedge \xi_2) = \sqrt{a^2} \, u$, $q^2 < 4$, and $N = \frac{1}{2}(q-1) |g|^{-1/2} [ q \, \xi_1
\widetilde{\otimes}  \xi_2   - 2\, \xi_1 \otimes \xi_1 - 2 \, \xi_2 \otimes \xi_2  ]= \varphi \, v \otimes v + \phi \, w \otimes w$. A straightforward calculation leads to:
\be
\hspace{-10mm} 
\varphi + \psi = \frac{1}{|g|}\,[ q \, \xi_1 \cdot \xi_2- \xi_1^2 - \xi_2^2] \, , \quad \varphi \, \psi = \frac{(4-q^2)}{4 |g|}[\xi_1^{2}\,  \xi_2^{2} - (\xi_1 \cdot \xi_2)^2] > 0  ,
\ee
and then $a^2 + \varphi \, \psi > 0$. Moreover, if $a=0$ ($q=0$) we obtain $\varphi + \psi <0$, and then lemma \ref{lemma-ricci2} implies that the Ricci tensor is algebraically special if, and only if, $\varphi = \psi$. And this condition holds when the Killing vectors $\xi_1$ and $\xi_2$ have the same modulus, $\xi_1^{2} =  \xi_2^{2}$, and $2 \, \xi_1 \cdot \xi_2 = q \, \xi_1^{2}$.
Thus, we obtain:
\begin{proposition}  \label{propo-ricci-7}
If a Bianchi type VII acts transitively, then the Ricci tensor is algebraically general if, and only if $N$ is algebraically general. Otherwise, when $N$ has two equal eigenvalues the Ricci tensor is:
\be
R = -2 a^2 g  \, .
\ee
\end{proposition}
Note that in the Bianchi type VII, if the Ricci is algebraically special, then the metric has a non-positive constant curvature. Then, $N$ admits an eigenplane that contains two equimodular Killing vectors. In type VII$_0$ these Killing vectors are also orthogonal, and the metric is flat.


And finally, we consider the cases when $N$ has rank three, that is, types VIII, and IX. Now $a=0$, and $N$ has non vanishing eigenvalues. Then from (\ref{ricci2}) we obtain: 
\begin{lemma} \label{lemma-ricci3}
In Bianchi types VIII and IX, $N = \sum \varphi_i \, u_i \otimes u_i $, the Ricci tensor is
\be \label{ricci-89} 
R= \sum \alpha_a \, u_a \otimes u_a  \, , \qquad \alpha_a = \frac12 [\varphi_a^2 -(\varphi_b - \varphi_c)^2] \, , \quad  a,b,c \not= \, .
\ee
\end{lemma}
Note that the Ricci eigenvectors are the eigenvectors of $N$.  From (\ref{ricci-89}) follows that the Ricci tensor is $R = \frac12 \varphi_1^2 \, g$ when $N$ has three equal eigenvalues, $\varphi_1 = \varphi_2 = \varphi_3$. This case is only compatible with signature 3 and it is forbidden in type VIII. If $N$ has two different eigenvalues, $\varphi_1 = \varphi_2 \not= \varphi_3 $, then the Ricci tensor $R$ has the same algebraic type, $\alpha_3 =\frac12 \varphi_3^2 > 0$,  $\alpha_1 = \alpha_2 = \frac12 \, \varphi_3 (2 \varphi_1 - \varphi_3)$; moreover $\alpha_1 + \alpha_3 = \varphi_1 \varphi_3$ is positive for signature 3 (type IX) and negative for signature 1 (type VIII). And there is another case where $N$ is algebraically general but the Ricci tensor has also two different eigenvalues: if $\varphi_2 + \varphi_3 = \varphi_1 \not= 2 \varphi_2$, then  $\alpha_1 =2 \varphi_2 \varphi_3$, $\alpha_2 = \alpha_3 =0$. Thus, we can state:
\begin{proposition}  \label{propo-ricci-8-9}
If a Bianchi type VIII (respectively, type IX) act transitively, then the Ricci tensor is algebraically special with two different eigenvalues if, and only if, one of the following conditions holds:
\begin{itemize}
\item[(i)] $N$ has also two different eigenvalues, $\varphi_1 = \varphi_2 \not= \varphi_3 $, and then:
\be
\hspace{-25mm}
R = (\alpha - \beta) u \otimes u + \beta  g \, ,  \quad \alpha > 0\, , \quad  \alpha + \beta < 0 \ ({\rm respectively},\ \alpha + \beta > 0) \, .
\ee
\item[(ii)] $N$ is algeraically general and $\varphi_2 + \varphi_3 = \varphi_1 \not= 2 \varphi_2$, and then:
\be
R = \alpha \, u \otimes u \, , \qquad \alpha < 0  \ ({\rm respectively},\ \alpha  > 0) \, . 
\ee
\end{itemize}
Moreover, in Bianchi type IX the Ricci tensor is proportional to the metric, $3 R =r g$, $r >0$,  if, and only if, $N$ is also proportional to the metric.
\end{proposition}


\section{Algebraically special case}
\label{A-AS}

Let us suppose that the Ricci tensor of a homogeneous three-dimensional Riemannian space is algebraically special. We have:
\begin{equation} \label{riccisp}
R = (\alpha - \beta) u \otimes u + \beta g \, , \qquad  \dif \alpha =0= \dif \beta \, .
\end{equation}
Then, Bianchi identities $2\, \nabla\! \cdot \! R = \dif r$, imply that $u$
is geodesic and expansion free. Moreover, two scalars $\sigma$, $\Omega$
exist (shear and rotation scalars) such that
\begin{equation}  \label{special1}
\nabla u = \sigma (v \otimes v - w \otimes w) + \Omega(v
\wedge w), \qquad  \dif \sigma =0, \quad  \dif \Omega =0 \, ,
\end{equation}
where $\{u, v , w \}$ is an orthonormal oriented frame. Also, a 1-form ${\cal C}$ exists such that
\begin{equation} \label{special2}
\nabla v = (- \sigma v + \Omega w) \otimes u + {\cal C} \otimes
w, \quad \nabla w = ( \sigma w - \Omega v) \otimes u -
{\cal C} \otimes v \, .
\end{equation}
The second structure equations for this frame and a Ricci like
(\ref{riccisp}) lead to
\begin{equation}
\Omega^2 - \sigma^2 = \frac{\alpha}{2} \, , \qquad  \sigma \, {\cal C} \wedge w = \sigma \, {\cal C} \wedge v = 0   \, , \qquad 
 \dif  {\cal C} = \beta \, \omega  \wedge v \, . \label{sp4}
\end{equation}
Now, two different cases arise. If $\sigma =0$, that is, when a G$_4$ exists, then the frame $\{ v,
\omega \}$ is not unique, and equations above just say that $\Omega^2 =
\frac{\alpha}{2}$ and $\dif {\cal C} = \beta \, \omega \wedge v$. Note that $\alpha = 0$ is forbidden when $\alpha + \beta =0$ since it implies $R=0$. Thus, we obtain:
\begin{proposition} \label{propo-sigma0}
A three-dimensional Riemannian metric that admits a G$_4$ as maximal group of isometries has an algebraically special Ricci tensor, $R = (\alpha-\beta) u \otimes u + \beta g$, $\alpha \neq \beta$, with a non-negative simple eigenvalue. More precisely, $\alpha = 2 \Omega^2 \geqslant 0$, where $\Omega$ is the rotation scalar of the simple eigenvector. When $\alpha + \beta = 0$, then $\alpha > 0$ ($\Omega \not=0$).
\end{proposition}

If $\sigma \neq 0$, an invariant frame exists and so the maximal
dimension of the isometry group is three. Equations (\ref{sp4}) lead to ${\cal C}= 0 , \ \Omega^2 - \sigma^2 = \frac{\alpha}{2},  \ \beta =0$. Thus, we obtain:
\begin{proposition} \label{propo-sigmanot0}
A three-dimensional Riemannian metric that admits a G$_3$ as maximal group of isometries and with an algebraically special Ricci tensor, $R = (\alpha-\beta) u \otimes u + \beta g$, $\alpha \neq \beta$, has a zero double eigenvalue, $\beta =0$, and the simple eigenvalue is $\alpha = 2 (\Omega^2 - \sigma^2)$, where $\Omega$ and $\sigma$ are the rotation and the shear scalars of the simple eigenvector.
\end{proposition}

Now we want to analyze the Bianchi-Behr types that are compatible with an algebraically special Ricci tensor in each one of the two possible situations, namely, $\sigma = 0$ and $\sigma \not=0$. In \ref{A-Ricci} we have shown (propositions \ref{propo-ricci-2} and \ref{propo-ricci-3}) that the Ricci tensor of types II and III is algebraically special and the simple eigenvalue is proportional to a Killing vector. Thus, we obtain:
\begin{proposition} \label{propo-sigma-2-3} 
If the Bianchi types II or III act transitively, then the Ricci tensor is algebraically special and the simple eigenvector is shear-free. 
\end{proposition}

On the other hand, the transitive action of particular cases of the Bianchi-Behr types VI$_0$, $VIII$ and $IX$ also lead to an algebraically special Ricci tensor. One case occurs in types $VIII$ and $IX$ when $N$ has two equal eigenvalues (see proposition \ref{propo-ricci-8-9}). Then $a =0$, and substituting $Z=N = (\varphi - \psi) u \otimes u + \psi g$ in the invariant condition (\ref{nablaZ}), we obtain that the shear of the simple eigenvector $u$ vanishes. 

The other case occurs in types $VIII$ and $IX$ when the eigenvalues of $N$ fulfill $\varphi_2 + \varphi_3 = \varphi_1 \not= 2 \varphi_2$ (see proposition \ref{propo-ricci-8-9}). A similar constraint of the eigenvalues occurs in type VI$_0$ when $\tr N = \varphi_2 + \varphi_3 = 0$ (see proposition \ref{propo-ricci-6}). Again $a =0$, and substituting $Z=N = \sum \varphi_i \, u_i \otimes u_i $ in the invariant condition (\ref{nablaZ}), we obtain that the shear and the rotation scalars of the simple eigenvector $u=u_1$ of the Ricci tensor are, respectively, 
\be
\sigma = \frac12 (\varphi_3 - \varphi_2) \, , \qquad  \Omega = - \frac12 (\varphi_3 + \varphi_2) \, .
\ee
Note that $\varphi_2 + \varphi_3 = \varphi_1 =0$, and then $\Omega=0$, is only compatible with an $N$ of rank two, that is, type VI$_0$. Thus, in the last two paragraphs we have shown:
\begin{proposition} \label{propo-sigma-6-8-9} 
(i) If the Bianchi-Behr types VIII or IX act transitively and the Ricci tensor is algebraically special, then the simple eigenvector of the Ricci tensor $u$ is shear-free if, and only if, the structure tensor $N$ has two equal eigenvalues. When $u$ has a non-vanishing shear, then it has a non-vanishing rotation. \\
(ii) If the Bianchi-Behr type VI$_0$ acts transitively and the Ricci tensor is algebraically special then the simple eigenvector of the Ricci tensor has a non-vanishing shear and it is rotation-free.
\end{proposition}
%


We know (see \cite{bonacoll1, bonacoll2} and our results in this appendix) that, when the Ricci tensor is algebraically special, $R = (\alpha - \beta) u \otimes u + \beta g$, the shear and the rotation of the simple eigenvector $u$ allow us to pick out several Bianchi-Behr types. In order to obtain an IDEAL labeling of these cases we must obtain these kinematic coefficients as explicit concomitants of the Ricci tensor. 

If $h = g - u \otimes u$ is the projector on the two-plane orthogonal to $u$ and $\{u, v , w \}$ is an orthonormal oriented frame, 
we have:
\be
\nabla u =  u \otimes a + \frac12 \theta h + \sigma (v \otimes v - w \otimes w) + \Omega(v
\wedge w) \, . 
\ee
where $a$ is the acceleration, $\theta$ the expansion, $\sigma$ the shear scalar and $\Omega$ the rotation scalar. It is worth remarking that in homogeneous spaces the Ricci eigenvalues are constant and then the Bianchi identities implies $a=0$ and $\theta=0$, and we recover expression (\ref{special1}). Nevertheless, now we look for expressions for $\sigma$ and $\Omega$ without any a priory hypothesis of homogeneity. 

From the expression of the Ricci tensor we can obtain $\nabla R \cdot R$, a tensorial expression that contain $\nabla u$. Then we can build a two-tensor by contracting two indexes with the volume element and we can remove the terms containing $a$ and $\theta$ by contracting with the projector $h$ and obtaining the symmetric part. Finally, one obtains:
\be \label{sigmaOmega}
D_{ij} \equiv (\nabla R \cdot R)^{k l m} \eta_{m l (i}\, h_{j)k} \, , \qquad D = (\alpha-\beta)^2 (\sigma \, v \widetilde{\otimes} w - \Omega \, h) \, 
\ee
where $h$ can be obtained from the Ricci tensor as $h \equiv [2 \, S^3 - (4 s/3) \, S ]/t$.
%



\end{document}